\documentclass[3p,times,procedia]{elsarticle}
\usepackage{ecrc}
\volume{00}
\firstpage{1}
\journalname{Procedia Computer Science}
\runauth{}
\jid{procs}
\jnltitlelogo{Procedia Computer Science}
\CopyrightLine{2011}{Published by Elsevier Ltd.}
\usepackage{amssymb}
\usepackage{graphicx}
\usepackage{subfigure}
\usepackage{multirow}
\usepackage{amsmath}
\usepackage{algpseudocode}
\usepackage{algorithm}
\usepackage{listings}
\usepackage[figuresright]{rotating}
\begin{document}
\begin{frontmatter}
\dochead{4th International Conference on Ambient Systems,\\ Networks and Technologies (ANT), 2013}

\title{M-ATTEMPT: A New Energy-Efficient Routing\\ Protocol for Wireless Body Area Sensor Networks}

\author{N. Javaid$^{\ddag}$, Z. Abbas$^{\ddag}$, M. S. Fareed$^{\ddag}$, Z. A. Khan$^{\$}$, N. Alrajeh$^{\sharp}$}

\address{$^{\ddag}$COMSATS Institute of Information Technology, Islamabad, Pakistan. \\
        $^{\$}$Faculty of Engineering, Dalhousie University, Halifax, Canada.\\
        $^{\sharp}$B.M.T., C.A.M.S, King Saud University, Riyadh, Saudi Arabia.\\
}

\begin{abstract}
In this paper, we propose a new routing protocol for heterogeneous Wireless Body Area Sensor Networks (WBASNs); Mobility-supporting Adaptive Threshold-based Thermal-aware Energy-efficient Multi-hop ProTocol (M-ATTEMPT). A prototype is defined for employing heterogeneous sensors on human body. Direct communication is used for real-time traffic (critical data) or on-demand data while Multi-hop communication is used for normal data delivery. One of the prime challenges in WBASNs is sensing of the heat generated by the implanted sensor nodes. The proposed routing algorithm is thermal-aware which senses the link Hot-spot and routes the data away from these links. Continuous mobility of human body causes disconnection between previous established links. So, mobility support and energy-management is introduced to overcome the problem. Linear Programming (LP) model for maximum information extraction and minimum energy consumption is presented in this study. MATLAB simulations of proposed routing algorithm are performed for lifetime and successful packet delivery in comparison with Multi-hop communication. The results show that the proposed routing algorithm has less energy consumption and more reliable as compared to Multi-hop communication.
\end{abstract}


\end{frontmatter}

\section{Introduction}

Patient monitoring is emerging as an important application of embedded sensors network. Many wireless sensors are implanted in/on the patient body. These tiny wireless sensors make Wireless Body Area Sensor Networks (WBASNs). A WBASNs can observe physiological conditions of patient under supervision, and can provide us real-time feedback. Through WBASN a patient is constantly monitored, and in case of some critical situation an immediate action should be required. These sensors can collect the physiological data and then send to physician in a hospital through Metropolitan Area Network (MAN) or Local Area Network (LAN). Where, diagnosis from received information is performed and on that base decisions are taken. WBASNs are used for medical and non medical applications. The wireless sensor nodes used in WBASNs are tiny, light-weight and of limited power. These sensor nodes have different energy levels and generate different size of data while the Wireless Sensor Networks (WSNs) nodes almost have same level of energy and data rate. Thus, employing routing algorithm of WSN can not support WBASNs sensor nodes. The selection of WBASNs routing algorithms should support the heterogeneous sensors network.

In [1], Single-hop communication is used between sensor nodes and sink node. To overcome the problem of topological partitioning due to constant human body movement and ultra short Radio Frequency (RF) transmission range. Sue \textit{et al.} [2] used Multi-hop communication for communication between sink and root nods. However, In direct communication increase in temperature of sensor nodes may affect human body tissues. He also discussed that storage delay (due to topological disconnections) and congestion delay increase overall delay in Multi-hop communication that can not be helpful for emergency services and makes Multi-hop communication not a best choice.

In this paper, a prototype for placing heterogenous sensor nodes on human body is presented. High data rate nodes are placed on less mobile places on human body. Mobility of human body cause disconnection between previous established links. It takes time to establish new connection to forward data and causes delay. Delay is not supportive in real-time applications. To beat delay and overcome problem of disconnection. We used energy management in our proposed routing protocol. By using energy management sensor nodes increase their transmission range and directly communicate with sink node for critical data delivery. For normal data delivery Multi-hop communication is used.

\section{Background and Motivation}

The rising temperature of implanted sensor nodes due to communication radiations and circuitry power consumption can affect the human body. In [3], authors use thermal-aware routing to minimize the effect of rising temperature of implanted sensor nodes.  Quwaider \textit{et al.} [4] used Single-hop communication and increased transmission range of sensors to overcome problem of partitioning.

Environment Adaptive Routing (EAR) algorithm [5] defines different communication cost for heterogeneous WBASNs devices. However, the Single-hop communication and proactive routing are not suitable choices for WBASNs. Multi-hop communication is suitable for normal packet delivery and Single-hop is only used for emergency services due to high transmission cost. However, here use of Hello messages after a regular interval results in high energy consumption.

In [6], Wireless Autonomous Spanning Tree Protocol (WASP) is defined to achieve low delay and increased network reliability for WBASNs. In WASP-scheme a message is disseminated to update parent nodes with information of its child nodes. However, power balancing issue is not discussed. Annur \textit{et al.} [7] applied tree algorithm with prioritization for WBASNs. Where, a channel is dedicated for emergency data delivery and normal data transmission is lagged until the successful delivery of critical data. However, the dedicated channel results in loss of available resources.

In this study, we present a routing protocol which works better in heterogeneous as well as homogeneous network. The nodes are placed around the sink in descending order of their data rate. A priority mechanism is used where low data rate sensors can not forward their data until all emergency data is transferred. To overcome delay, critical data is sent directly to the sink node and normal data is sent through Multi-hop communication. Our proposed model is discussed in next section.

\section{System Model}

\begin{figure*}[ht]
  \centering
 \subfigure{\includegraphics[height=3 cm, width=10 cm]{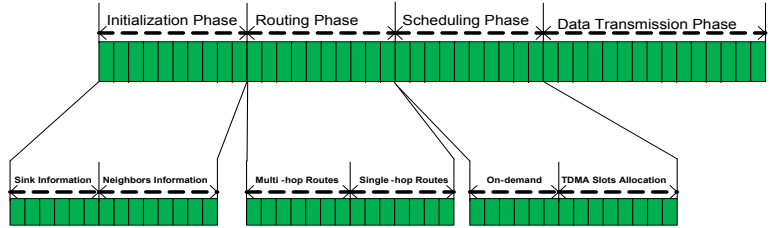}}
  \caption{Sequence of Phases in Each Round}\label{fig2}
\end{figure*}

In our proposed model, sink is placed at center of the human body. Since Wireless Body Area Sensor Networks (WBASNs) are heterogeneous networks, then placement of nodes on human body is an issue. This issue is resolved by placing nodes in descending order of their data rate with respect to sink. Thus, the nodes with high data rate send data directly to the sink node, and can easily forward the received data from low data rate sensors. Problems analyzed in previous work are set in following manner: (1) Single-hop communication is used for emergency services and on-demand data, (2) for normal data delivery Multi-hop communication is used, (3) path with less hop count is selected for network life time extension in Multi-hop communication. Fig. \ref{fig2} depicts the phases of proposed routing protocol with above mentioned features. Complete description of the proposed routing protocol is provided in next section.

\subsection{Initialization Phase}
In initialization phase, all nodes broadcast Hello messages. This Hello message contains neighbors information and distance of sink nodes in form of hop-counts. In this way, all nodes are updated with their neighbors, sink position and available routes to the sink node. Route computation for data delivery to sink node of the proposed routing algorithm is discussed in next subsection.

\subsection{Routing Phase}
In this phase, routes with fewer hops to sink are selected from available routes. We suppose nodes have information of all nodes and sink's position. So, selected routes are steadfast and consume less energy. Emergency services are also defined in proposed routing protocol. In critical scenarios, all processes are lagged until critical data is successful received by sink node. In case of emergency, all the implanted nodes on the body can communicate directly with the base station. Moreover, all sensor nodes can communicate directly with the sink node when demand is arrived from sink. In direct communication, delay is much lower as compared to Multi-hop communication, because in Multi-hop communication, each intermediate node receives, processes and then sends data to next node. The reception, processing and transmission of received data on each intermediate node takes time which causes delay. Sometime, this delay is also increased due to congestion and becomes unacceptable in some critical scenarios. So, single-hop communication is used to minimize this delay. We calculate energy consumed in Single-hop communication as:

\begin{align}
  E_{S-HOP}=E_{transmit}
  \end{align}
 where, the transmission energy is calculated as:
\begin{align}
  E_{transmit}= E_{elec}+ E_{amp}
  \end{align}

where, $E_{elec}$ is the energy consumed for processing data and $E_{amp}$ is energy consumed by transmit amplifier. We suppose a linear network in which all nodes are implanted at equal distance from each other. To transmit $b$ bits up to $n$ hops the transmission energy is given as:

where $d^{2}$ is the energy loss due to the transmission.

\begin{align}
    E_{transmit}= n\times b(E_{elec}+ E_{amp})\times d^{2}
    \end{align}
\begin{figure*}[ht]
  \centering
 \subfigure{\includegraphics[height=5 cm, width=10 cm]{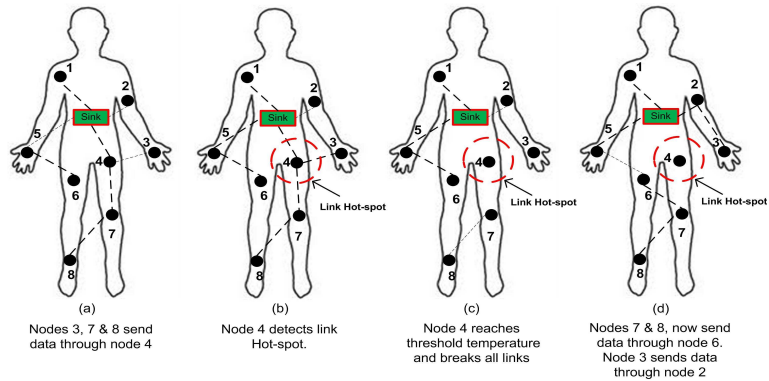}}
  \caption{Link Hot-spot Detection}\label{fig3}
\end{figure*}

In Single-hop/Multi-hop traffic control algorithm 1, if a node senses emergency or on-demand data, node uses Single-hop communication. In Single-hop communication, the sensor node uses full power of battery to send its data. If normal data is received, the Multi-hop communication is used to send data to sink node, which results in less energy consumption. More energy is required to send data at greater distances. Thus, in Multi-hop communication, energy consumption is very less.

Energy preservation is a prime consideration in WBASNs, as the deployed sensor nodes  have limited energy sources. So, deployed nodes need reasonable use of battery for extended network life time. To calculate the energy consumption during a Multi-hop communication we assume a linear network in which all nodes are deployed at equal distance from each other. The energy consumption during Multi-hop communication can be computed using the following equations:

\begin{align}
  E_{M-HOP}= E_{transmit}+E _{received}
  \end{align}

 where, $E _{received}$ is the energy consumed for receiving data. If we are transmitting $b$-bits to a distance of $n$-hops the transmission energy will be $n\times b\times E_{transmit}$ and receiving energy will be $(n-1)b\times E _{received}$. Since the first node transmits only and intermediate nodes receive $n$-bits and then transmit these received bits. So, the energy consumed for Multi-hop is:

\begin{align}
E_{M-HOP} = n\times b\times E_{transmit}+ (n-1)b\times E _{received}
\end{align}

From (2) and (6), taking $E _{received}=E_{elec}$ since the receiving energy is equal to energy consumed to process received data we obtain

\begin{align}
E_{M-HOP} = n\times b\times (E_{elec}+ E_{amp}\times d^{2}) + (n-1)b\times E_{elec}
\end{align}

\begin{align}
E_{M-HOP}= n\times b\times E_{elec}+ n\times b\times E_{amp}\times d^{2}+nb\times E_{elec}-b\times E_{elec}
\end{align}

\begin{align}
E_{M-HOP}= [2\times n\times b\times E_{elec}+ nb\times E_{amp}\times d ^{2}-b\times E_{elec}]
\end{align}

When we are dealing with wireless communication around the human body, effects of these sensors on human body can also be taken into consideration. The most important considered factor for this purpose is Specific Absorption Rate (SAR) and heating effects of the implanted sensor nodes on human body. Taking SAR and heating effects of sensors on human body in mind, we considered and implemented link hot-spot detection method. Here, nodes implanted closer to sink node are forwarding data of their follower nodes. Whenever, a temperature threshold is reached, a node breaks its link with its neighbor for few rounds. As temperature returns to normal, it re-establishes the original route. However, if a sensor node receives a data packet and reaches its temperature threshold it returns packet to previous node. And previous node mark this link as Hot-spot as shown in Fig.  \ref{fig3}.

\begin{algorithm}[H]
\caption{: ATTEMPT Algorithm}
\begin{algorithmic}[1]
\State \textbf{Routing Phase}
            \If{( route\_ 1 $<$ route\_  2 )}
                \State route\_  1 = selected route
              \Else
               \State route\_  2 = selected route
               \If{( route 2 $<$ route\_  1 )}
                \State route\_  2 = selected route
              \Else
               \State route\_  1 = selected route
               \If{( route \_ 1 = route \_ 2 )}
               \State $E_{hop-count} \gets$ Energy consumption for a route
               \If{($E_{hop-count} \_1 < E_{hop-count}$ \_2 )}
                \State route\_  1 = selected route
              \Else
               \State route\_  2 = selected route
                          \EndIf
                          \EndIf
                          \EndIf
                             \EndIf
                          \end{algorithmic}
                           \end{algorithm}

\begin{figure*}[ht]
  \centering
 \subfigure{\includegraphics[height=3 cm, width=12 cm]{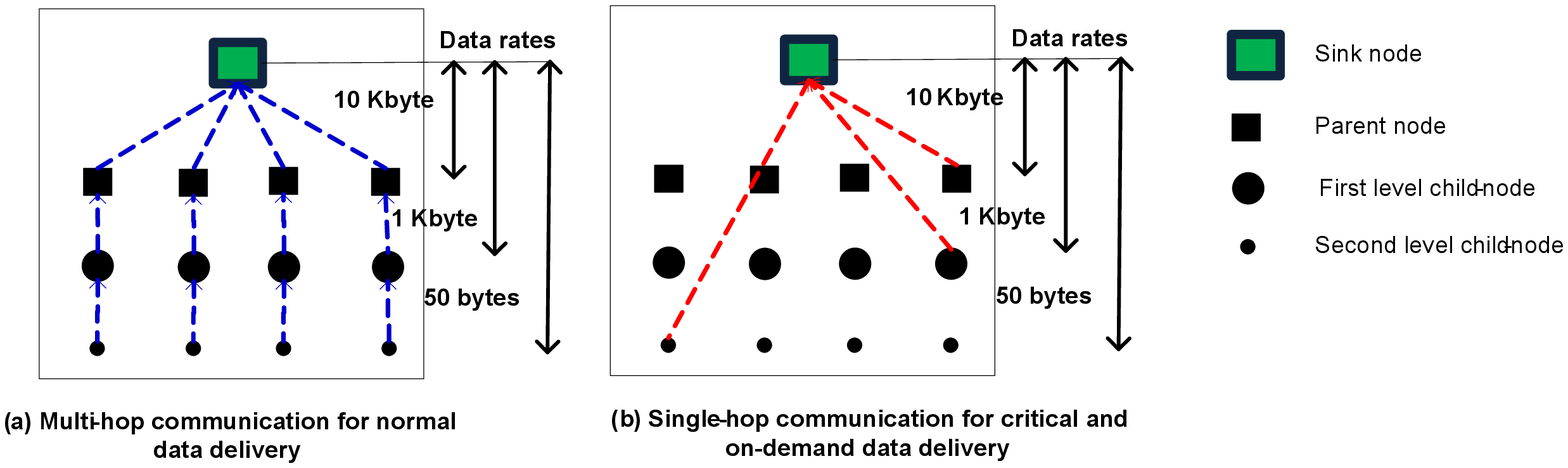}}
  \caption{Energy Management for Single-hop and Multi-hop Communication}\label{fig4}
\end{figure*}

ATTEMPT routing is discussed in Algorithm 2. If two routes are available then route with less hop-counts is selected. If two routes have same hop-count, then route with less energy consumption to the sink is selected. Single-hop and Multi-hop communication of root node with sink is shown in Fig. \ref{fig4}.

\subsection{Scheduling Phase}
After route selection the sink node creates Time Division Multiple Access (TDMA) schedule for communication between sink node and root nodes. Sink node allocate time-slots to nodes. Nodes can communicate to sink node in assigned time slot for normal data delivery.

\subsection{Data Transmission Phase}
Once the time slots are allocated to root nodes, root nodes send their data to sink node in assigned time slot. After that sink node will receive data, and will take some time to aggregate the received data. 	

\section{Mobility Support in ATTEMPT}
To introduce mobility support in ATTEMPT, we defined a prototype for placing nodes on human body for Mobile-ATTEMPT (M-ATTEMPT). Nodes with high data rates are placed at less mobile places on human body. These high data rate nodes are parent nodes and are directly connected to sink. Parent nodes have $10$$J$ energy and generate $10$ $Kbytes$ of data. The nodes directly connected to parent nodes are first level child-nodes with $5$$J$ energy and generate $1$ $Kbyte$ of data. The nodes connected to first level nodes are second level child-nodes with $1$$J$ energy and generate $50$ $bytes$ of data. Parent nodes, first level child-nodes and second level child nodes placed on human body and their respective topology is shown in Fig. \ref{fig4}.

\begin{figure*}[ht]
  \centering
 \subfigure{\includegraphics[height=8 cm, width=12 cm]{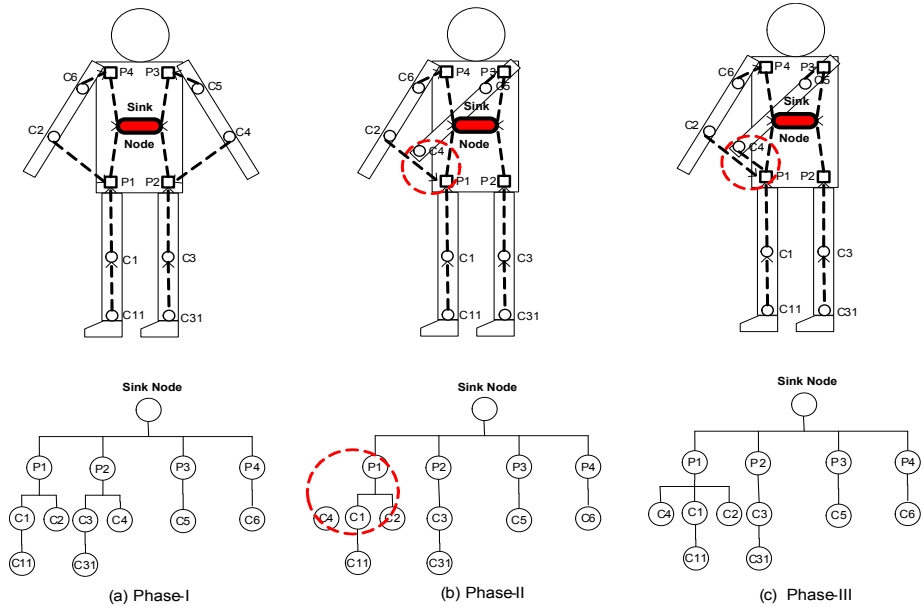}}
  \caption{Link Establishment and Link Breakage Due to Mobility of Human Body}\label{fig6}
\end{figure*}

\subsection{Invitation Phase}
In this phase, we will discuss how our proposed routing protocol support mobility. if a node changes its position during a round, nodes have to pay lot of energy to maintain this established link. Consider an example where, first level child node $C4$ is disconnected from its parent node $P2$ and entered in communication range of parent node $P1$. Now $C4$ will send joint-request to parent node $P1$. Parent node will check its parent child list if number of child nodes are less than $3$. Then, parent node $P1$ will accept joint-request and register $C4$ as a child node as depicted in Fig. \ref{fig6}.

\section{Simulation Results}
We performed series of simulations to compare performance of our proposed routing protocol with  Multi-hop. We used MATLAB as a simulator to analyze the performance of proposed routing protocol. We took network size of $5m$ $\times$ $5m$ in which $10$ nodes are randomly distributed and sink node is placed in the center of the network. 5000 number of rounds taken and nodes with initial energy of nodes $0.5J$ and radio range of $10m$

\begin{figure}[h!]
\begin{center}
\includegraphics[scale=0.28]{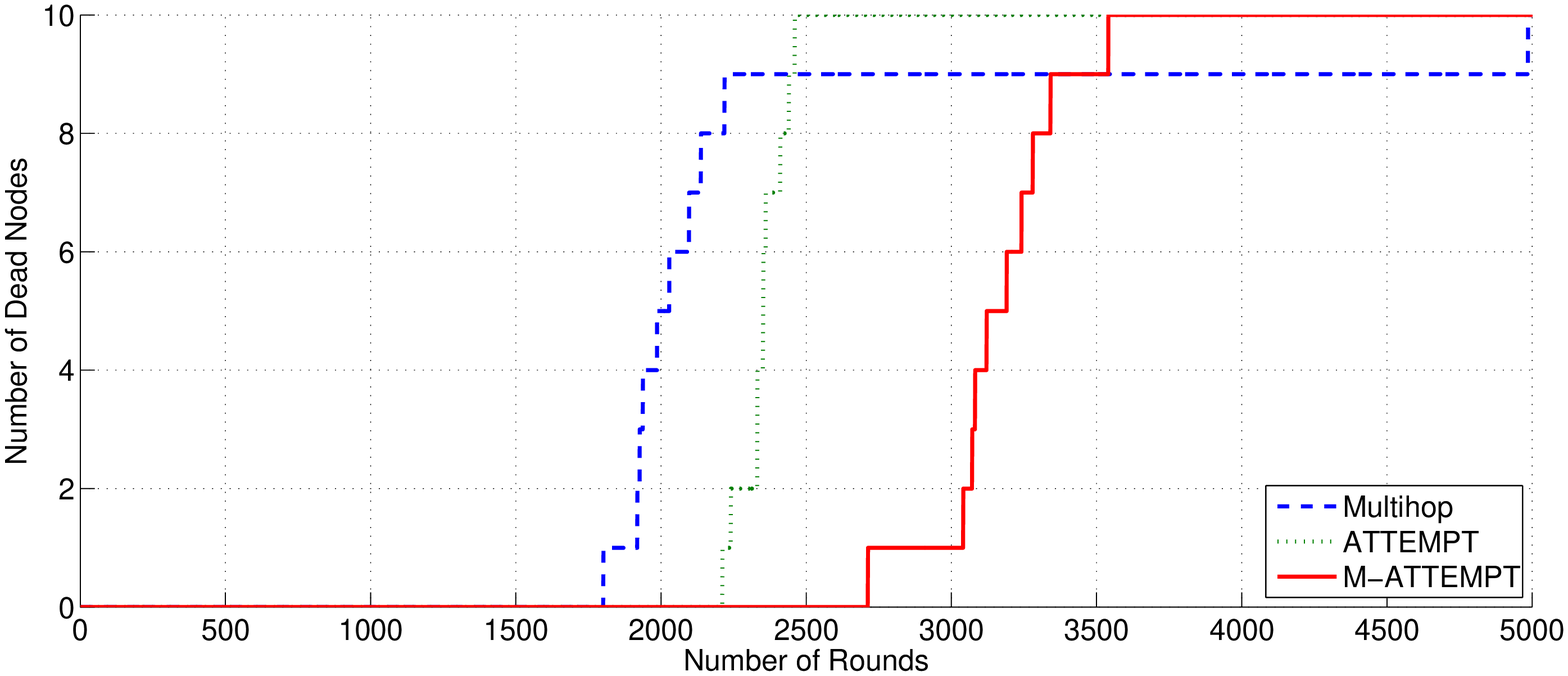}
\caption{Number of Dead Nodes Over Time }\label{fig8}
\end{center}
\end{figure}

The time in which all nodes in network are alive is called the stability period of network. M-ATTEMPT in sense of network life time performs best as long as compared to Multi-hop communication and has almost greater lifetime as compared to ATTEMPT. Number of dead nodes are presented as function of rounds. Round is time required to establish a network, here probability for network establishment is kept to be 10\%. Where after this value node can be selected as CH. So, taking this parameter first node dies out at 2700 round for M-ATTEMPT as compared to others. M-ATTEMPT has better stability period as compared to Multi-hop routing or ATTEMPT as depicted in Fig. \ref{fig8}.

 \begin{figure}[h!]
\begin{center}
\includegraphics[scale=0.28]{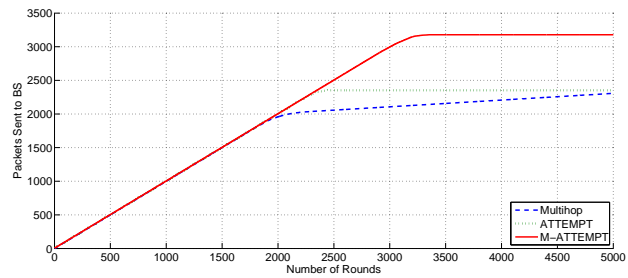}
\caption{Throughput }\label{fig9}
\end{center}
\end{figure}

A Multi-hop routing is the best choice for WBASNs. Here, in our proposed routing protocol single-hop and multi-hop both concepts are being used. So, Throughput of both techniques implemented with mobility support is experimentally investigated, by measuring the successful packets delivery at BS. Fig. \ref{fig9} depicts that throughput of M-ATTEMPT is almost better in stable and unstable region than ATTEMPT and Multi-hop communication. ATTEMPT sends threshold data and periodic data, so its throughput is estimated from this data. In multi hop communication only periodic data is received at sink node. While, M-ATTEMPT which uses Single-hop and Multi-hop communication with mobility management performs better as compared to Multi-hop and ATTEMPT. In case, if link is broken because of mobility, M-ATTEMPT establishes link with another node by checking hop counts or on minimum energy usage basis. So, no loss of packets occur in case of link breakage.

 \begin{figure}[h!]
\begin{center}
\includegraphics[scale=0.28]{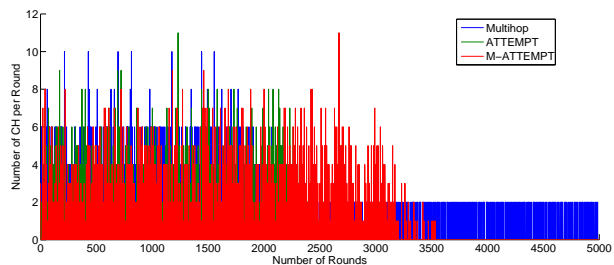}
\caption{Number of Cluster Heads Per Round }\label{fig10}
\end{center}
\end{figure}
  \begin{figure}[h!]
\begin{center}
\includegraphics[scale=0.28]{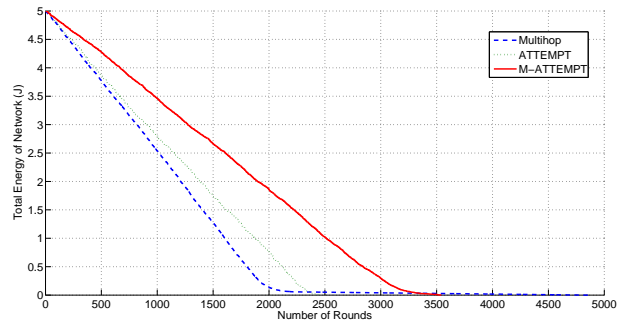}
\caption{Total Energy of Network over Varying Rounds}\label{fig11}
\end{center}
\end{figure}

Fig. \ref{fig10} shows number of selected CH in each round for all three protocols. Here, M-ATTEMPT shows maximum CH selection in some rounds, as in some rounds there are still chances for CH selection for other nodes. Total energy consumption of M-ATTEMPT, ATTEMPT and multi hop communication is depicted in  Fig. \ref{fig11}. M-ATTEMPT utilizes maximum energy as nodes start dying after reaching maximum rounds as compared to ATTEMPT and Multi-hop communication.

Finally, a comparison in terms of percentage between all protocols is discussed. Results of all three techniques clarifies that our proposed protocol's stability period is 20\% greater as compared to Multi-hop and 11\% greater in comparison with ATTEMPT over changing rounds. This is because that our proposed protocol uses  Single-hop and Multi-hop with mobility management technique. When comparison between number of successfully received packets at sink is performed between all three techniques, then M-ATTEMPT shows a 29\% better results as compared to Multi-hop and 12.5\% with ATTEMPT. M-ATTEMPT as compared to others has 29.5\% better network life time over varying number of rounds.

\section{Conclusion}
In this work, we presented an energy efficient routing algorithm for heterogeneous WBASNs. For real-time and on-demand data traffic root node directly communicates with sink node and for normal data delivery Multi-hop communication is used. Our proposed routing protocol supports mobility of human body with energy management. The proposed routing algorithm is thermal-aware which senses the link Hot-spot and routes the data away from these links. After selection of routes sink node creates TDMA schedule for communication between sink node and root nodes for normal data delivery using multihop communication. MATLAB simulations of proposed routing algorithm are performed for lifetime and packet delivery ratio in comparison with Multi-hop communication. Topology and placement of nodes is described with Single-hop and Multi-hop communication scenarios. The results show that proposed routing algorithm has less energy consumption and more reliable in sense of packet delivery as compared to Multi-hop communication.


\end{document}